# How do 'technical' design-choices made when building algorithmic decision-making tools for criminal justice authorities create constitutional dangers?

## by Karen Yeung[α] and Adam Harkens[β]

## Part I

### 1.    Introduction: the algorithmic turn in criminal justice decision-making

Automated, internet-enabled digital decision tools, particularly those that utilise some form of machine learning ('ML'), are widely touted as offering transformative 'solutions in government'.[1] Also called 'narrow' or 'task-specific' artificial intelligence ('AI'), ML is a computational technique that uses algorithms to find patterns and correlations in large datasets, 'learning' from past experience to create a mathematical model that can generate predictions when applied to unseen data. These models may, in turn, be embedded into a digital tool to inform, or even to automate, organisational decisions. The variety of ML-based algorithmic systems now being rapidly taken-up the public sector within EU Member States is showcased in the 'AI Watch' EU Joint Research Centre's 2020 Report,[2] with public sector adoption continuing apace.[3] Examples range from the use of automated image recognition technologies to detect whether agricultural grasslands have been mowed in Estonia, to the automation of welfare benefit decisions in Sweden, and the delivery of personalised social services to the unemployed in Poland.[4] British criminal justice authorities are also rapidly embracing these tools, incentivised by the government's Police Transformation Fund (and its predecessor) from which £720 million was allocated to projects aimed at transforming how police use technology between 2015-2020 alone.[5] The Home Office claims these technologies will 'deliver cash savings, as well as improving efficiency by, for example, freeing

---

[α] Professorial Fellow in Law, Ethics and Informatics, Birmingham Law School and School of Computer Science, University of Birmingham. We gratefully acknowledge funding support form VW Stiftung, Grant No: 19-0087 (2019-2023) and for helpful feedback by Emma Ahmed-Rengers (particularly in comparing conventional statistics with ML approaches), Reuben Binns, Mireille Hildebrandt, Tobias Krafft, Winston Maxwell, Leandro Minku, Johannes Schmees, Georg Wenzelberger. Karen Yeung drafted the text and devised the analytical framework, argument, paper structure and narrative. Adam Harkens undertook the in-depth case-studies and background research to the legal, scholarly and contextual detail supporting Yeung's arguments and acted as a critical sounding board for her ideas. An earlier version was presented by Karen Yeung to the Norwegian Association for Computers and the Law, The Knut Selmer Memorial Lecture, 23 November 2020 (Oslo).
[β] Post-doctoral Research Fellow, Birmingham Law School.

[1] Government Technology Innovation Strategy (10 June 2019) *https://www.gov.uk/government/publications/the-government-technology-innovation-strategy/the-government-technology-innovation-strategy*.
[2] Gianluca Misuraca and Colin van Noordt, "AI Watch: Artificial intelligence in public services," EU 30255 EN (Luxembourg: Publications Office of the European Union, 2020), 5; 41-47.
[3] Luca Tangi, Colin van Noordt, Marco Combetto, Dietmar Gattwinkel and Francesco Pignatelli, "AI Watch: European Landscape on the Use of Artificial Intelligence by the Public Sector" EUR 31088 EN (Luxembourg: Publications Office of the European Union, 2022), 37-38. They identified 686 tools in current use with more than half employing machine learning or automated reasoning and more than 50% of the tools identified are being implemented at the national level
[4] Misuraca and van Noordt, "AI Watch: Artificial intelligence in public services," 41-47.
[5] Home Office, "Police Funding for England & Wales: 2015-2021" (July 2020), Statistical Bulletin 16/20 *https://assets.publishing.service.gov.uk/government/uploads/system/uploads/attachment_data/file/900017/police-funding-england-and-wales-2015-to-2021-hosb1620.pdf*, 7**.** ML-based tools used by police forces range from geospatial systems that enable 'hotspot' and other forms of 'predictive policing' through to biometric identification systems such as live facial recognition technology. See A.G. Ferguson, *The Rise of Big Data Policing: Surveillance, Race, and the Future of Law Enforcement* (New York: NYU Press, 2017); Michael Veale, Sylvie Delacroix, Sofie Olhede, Christina Blacklaws and Sophia Adams-Bhatti, "Algorithms in the criminal justice system" (London: The Law Society of England and Wales, 2019).

up officers for frontline policing roles.'[6] Although global technology consultants make similarly celebratory claims that AI will generate better, more targeted, faster but less-costly decision-making,[7] the AI Watch 2020 report found 'little evidence of the social and economic impacts achieved so far,' observing that 'most digital transformations in the public sector seem to be guided by hopes and dreams, rather than confirmed by empirical evidence.'[8]

Academic scholarship across many disciplines also reflects a growing interest in public sector algorithmic tools, critically investigating how algorithms shape, coordinate and regulate behaviour and their implications for individuals, groups, public administration and political and public life more generally.[9] Prominent North American critics highlight how automated algorithmic tools, particularly those employed in public services including the criminal justice sector, deepen and reinforce existing economic and social inequalities[10], create pernicious feedback loops that operate in an opaque, unquestioned and unaccountable fashion[11] while fuelling a shift towards inequitable, undemocratic governance.[12] Legal scholars have emphasised the adverse implications arising from the take-up of algorithmic tools in public administration for 'rule of law values', such as transparency, fairness and accountability, reflecting on the role of administrative law[13] and human rights as 'ethical' constraints on AI systems[14] particularly when used in criminal justice contexts. Scholarship investigating the use of these tools in criminal justice contexts have traced the political and economic imperatives of the turn to so-called 'evidence-based' approaches to penal policy (including sentencing)[15] while highlighting problems of racial bias and gender discrimination (particularly in US contexts)[16], the problematic assumptions upon which they rest,[17] the paucity of evidence of their effectiveness[18], the expanding net, and

---

[6] 'Home Office awards over £100 million to police transformation projects' (Home Office Press Release, 1 August 2018), https://www.gov.uk/government/news/home-office-awards-over-100-million-to-police-transformation-projects.

[7] See for example InData Labs, "How artificial intelligence will change decision making" (December 2021) *shorturl.at/wzCX1*; Microsoft, "What is artificial intelligence?" https://indatalabs.com/blog/artificial-intelligence-decision-making#:~:text=AI%20decision%20making%20allows%20businesses,analyze%20large%20datasets%20without%20error.

[8] Misuraca and van Noordt, "AI Watch: Artificial intelligence in public services," 15.

[9] For example, Anneke Zuiderwijk, Yu-Che Chen and Fadi Salem, "Implications of the use of artificial intelligence in public governance: A systematic literature review and a research agenda" (2021) 38(3) *Government Information Quarterly*; (2021) 3 *Data and Policy*; Philip Alston, 'Report of the Special Rapporteur on Extreme Poverty and Human Rights" A/74/48037 (New York: United Nations, 2019). K. Yeung and M. Lodge (eds.), *Algorithmic Regulation* (Oxford: Oxford University Press, 2019); M. Hildebrandt, *Smart Technologies and the End(s) of Law* (Cheltenham: Edward Elgar, 2016)

[10] V. Eubanks, *Automating Inequality* (New York: St. Martin's Press, 2018), Ch.1.

[11] C. O'Neil, *Weapons of Math Destruction* (New York: Crown Books, 2016).

[12] K. Crawford, *Atlas of AI* (New Haven: Yale University Press, 2021).

[13] David Freeman Engstrom, Daniel E. Ho, Catherine M. Sharkey, Mariano-Florentino Cuellar, "Government by algorithm: Artificial intelligence in federal administrative agencies" (Report submitted to the Administrative Conference of the United States, 2020); Rebecca Williams, "Rethinking administrative law for algorithmic decision-making" (2021) 42(2) *Oxford Journal of Legal Studies* 468-494; Jennifer Cobbe, "Administrative Law and the Machines of Government: Judicial Review of Automated Public-Sector Decision-Making" (2019) 39(4) *Legal Studies* 63; Joe Tomlinson, "Justice in automated administration" (2020) 40(4) *Oxford Journal of Legal Studies* 708-736; Monika Zalnieriute, Lyria Bennett Moses and George Williams, "The rule of law and automation of government decision-making" [2020] 82(3) *Modern Law* Review 425-455; Katherine Freeman "Algorithmic Injustice: How the Wisconsin Supreme Court Failed to Protect Due Process Rights in State v Loomis" (2016) 18(5) *N.C.J.L & Tech* 75-106, 75.

[14] K. Yeung, A. Howes and G. Pogrebna, "AI governance by human rights-centred design, deliberation and oversight: An end to ethics-washing" in M. Dubber, F. Pasquale and S. Das (eds.) *The Oxford Handbook of AI Ethics* (OUP, 2019).

[15] B. Harcourt, *Against Prediction: Profiling, Policing, and Punishing in an Actuarial Age* (Chicago: University of Chicago Press, 2008).

[16] Julia Angwin, Jeff Larson, Surya Mattu, and Lauren Kirchner, 'Machine bias: There's software used across the country to predict future criminals. And it's biased against blacks" (ProPublica, 2016) at https://www.propublica.org/article/machine-bias-risk-assessments-in-criminal-sentencing; William Dietrich, Christina Mendoza and Tim Brennan, "COMPAS risk scales: Demonstrating accuracy equity and predictive parity (2016)" Northpointe Inc. https://go.volarisgroup.com/rs/430-MBX-989/images/ProPublica_Commentary_Final_070616.pdf;

[17] C. Barabas, "Beyond bias: 'Ethical AI' in criminal law" in M. Dubber, F. Pasquale and S. Das (eds.), *The Oxford Handbook of Ethics of AI* (OUP, 2020), and literature cited therein.

[18] Lyria Bennett Moses and Janet Chan, 'Algorithmic prediction in policing: assumptions, evaluation and accountability' (2018) 28(7) *Policing and Society*, 806-822.

intensification of the speed and scope of digital surveillance upon which they rely,[19] and their growing use to inform pre-emptive interventions aimed at combatting terrorism.[20] This two-part paper shares with this critical body of scholarship a foundational belief in the need to secure meaningful algorithmic accountability, taking as our point of departure Sarah Brayne's observation following her five year ethnographic study of 'big data policing' by the Los Angeles Police Department (LAPD) that:

> "Recognizing that data both shapes and is shaped by the social world helps us to understand that despite the argument made by proponents that big data policing is more objective and less discretionary than "human" decision-making, databases are populated by information collected as a result of human decisions, analyzed by algorithms created by human programmers, and implemented and deployed on the street by human officers following supervisors' orders and their own intuition."[21]

This paper critically investigates the legal and constitutional implications of the centrality of the human choices that are typically concealed behind a veneer of 'mechanical objectivity'[22] accompanying the algorithmic turn, building upon existing legal scholarship concerned with 'rule-of-law values' and administrative law doctrine. Yet our aim is to reach further and deeper, by demonstrating in more precise and concrete terms how seemingly 'technical' choices made by developers when building these tools have serious legal and constitutional implications, with a specific focus on the criminal justice context. The originality and significance of our paper lies in seeking to *integrate* public law principles and insights from public law scholarship with insights from principles and conventional practices of data scientists to demonstrate that technical design choices *directly* implicate constitutional principles and cannot be reduced to matters of technical computational know-how. Our analysis is primarily directed at lawyers, particularly those who believe that lawyers and legal theorists need not concern themselves with the 'technical' choices made by developers. We demonstrate that this belief is mistaken.

Our analysis proceeds across the two Parts as an integrated whole, although each Part may be read independently. Part I of this paper establishes the general claim that design-choices in algorithmic model-building to inform criminal justice decision-making directly implicate constitutional principles. Part II demonstrates in more precise and concrete terms how particular constitutional principles and legal obligations are directly implicated by specific 'technical' choices yet are typically ignored, drawing on three algorithmic decision-tools either in use or recently in use to illustrate our arguments. Although the analysis provided in Part II is dense and technical, it provides the foundation for our overarching argument – namely, that if algorithmic systems do not comply with public law norms and principles, they should not be used. While public lawyers constitute our primary target audience (who may wish to skip over our account of constitutional principles in Part I, section 3.2) we hope our paper is of interest to technical developers (who may wish to skip over our account of the technical dimensions of the algorithmic model-building in Part I, section 2) so that they might understand why public law principles, and the specific legal duties to which they give rise, cannot be ignored or partitioned off during algorithmic tool-development. Given that technical developers cannot be expected to understand constitutional principles and public law duties, we argue that they must collaborate closely with legal experts when deciding whether to employ decision-support tools for specific criminal justice purposes, and if justified, to ensure that they are configured in a manner that is demonstrably compliant with public law principles (including respect for human rights) throughout the tool-building process. If such compliance cannot be demonstrated, they should *not* be used.

---

[19] Elizabeth E. Joh, 'Policing by numbers: Big data and the fourth amendment' (2014) 89(1) *Washington Law Review*, A.G. Ferguson, *The Rise of Big Data Policing;* S. Brayne, *Predict & Surveil: Data, Discretion, And the Future of Policing* (New York: OUP, 2021); David Lyon, 'Surveillance, Snowden, and Big Data: Capacities, consequences and critique' (2014) *Big Data & Society*, 1-13; Mark Andrejevic, "Automating Surveillance" (2019) 17(1) *Surveillance and Society* 7-13.
[20] See Fahad Ahmed and Jeffrey Monahan 'From probabilities to possibilities: terrorism peace bonds, pre-emptive security and modulations of criminal law' (2020) 74(5) *Crime Law and Social Change,* 341-359 (and literature cited therein); L. Amoore, *The Politics of Possibility: Risk and Security Beyond Probability* (Durham: Duke University Press, 2013).
[21] Brayne, *Predict & Surveil: Data, Discretion, And the Future of Policing*, 138-139.
[22] T Porter, *Trust in Numbers: The Pursuit of Objectivity in Science and Public Life* (2nd edn., New Jersey: Princeton University Press, 2020).

Our methodological approach is unavoidably cross-disciplinary. In this Part I, we explain the primary objectives and reasoning processes of two quite distinct disciplinary perspectives – that of computer and data science on the one hand, and public law on the other. We argue that if algorithmic tools are intended to inform determinations by criminal justice authorities concerning how a given individual shall be treated, those tools must be constructed and designed in a manner that demonstrably conforms public law principles. Thirdly, we examine the constitutional acceptability of using algorithmic prediction models to inform determinations about how an individual should be treated by public authorities. Because these models rely on correlations between variables without an understanding of their underlying causal relationships, this substantially increases the dangers of relying on their statistical predictions to inform rights-critical decisions. We then highlight three constitutional principles of particular importance when building algorithmic tools that purport to assess the 'risk' posed by individuals to inform criminal justice decisions about how that individual should be treated (hereafter 'individual risk assessment tools' or 'i-RATS'): the rule of law (including the need for legal authorisation), respect for human rights generally, and the right to due process in particular, including more specific statutory and common law duties that can be understood as rooted in constitutional principle. We illustrate our arguments by way of example, using a familiar context within which algorithmic tools are now widely used in the USA: namely, decisions about the detention or release of an individual who has been arrested by police in England and Wales and taken into custody.[23] The final section concludes.

We begin with an account of algorithmic model-building, highlighting key differences between what we call 'conventional' statistical reasoning to create actuarial prediction models, and techniques rooted in data science methods, to explain how ML-based decision-support tools ('algorithmic tools') are created and intended to work. We draw attention to the 'abstraction decisions' of technical developers when creating an algorithmic prediction model intended to assist real-world decision-makers. This process conventionally entails detaching and excluding contextual considerations as irrelevant 'noise' from model-building, which we refer to as a 'contextual detachment mindset'. Secondly, we explain why decisions by criminal justice authorities are typically described as 'high stakes', highlighting the unique authority occupied by criminal justice officers which is ultimately derived from the general public and conferred upon them by law, unlike structural engineers and medical professionals whose authority to make 'safety-critical' decisions is rooted primarily in their professional expertise. Hence contexts in which criminal justice authorities make decisions about individuals are 'high stakes' because they are primarily 'rights-critical' rather than 'safety-critical'. Due to their unique coercive power conferred to enforce the criminal law on behalf of the state, decisions taken by criminal justice authorities are vulnerable to abuse and overreach and, if exercised arbitrarily, can produce serious injustice.[24] Recognising this, liberal democratic communities proceed on a foundational commitment to constitutionalism, insisting on institutional constraints that include constitutional principles, administrative law doctrine and human rights norms, all serving as important safeguards against the abuse of power.

## 2. Statistical foundations and 'technical' abstraction choices in algorithmic model-building

To understand why and how technical decisions taken during the algorithmic tool-building process directly implicate constitutional principles, a basic understanding of the mathematical logic and technical mechanisms underpinning algorithmic tools and the tool design-and-build process is required. Although statistical tools have been used to evaluate the 'riskiness' of individuals by criminal justice officials since the late 1920s,[25] more recent algorithmic tools that utilise ML have attracted greater scholarly interest and concern.[26] Fundamental doubts remain concerning the epistemic validity and legitimacy of using statistical tools to inform the state-mandated treatment of individuals based on predictions about future events that are, by definition, uncertain, particularly in the case of individual proclivities, such as estimates of 'dangerousness' or 'recidivism risk.'[27] Although these concerns are beyond the scope of our inquiry, our analysis reinforces these doubts. Although ML techniques can significantly enhance prediction-accuracy compared to more conventional statistical tools, their enhanced power and complexity is a double-edged sword when used to create decision-support tools to inform 'rights-critical' decisions, magnifying the dangers of error, injustice and the abuse of power in a variety of ways. To understand why, it is first necessary to understand the difference between conventional statistical methods and ML techniques that rely on advanced statistical techniques utilising computational power to automate the task of data analysis.[28]

---

[23] See section 3 below.
[24] These include the wrongful conviction of the the 'Guildford Four' on the basis of false confessions elicited by police coercion: See *R. v. Maguire and others* (1991) 94 Crim. App. R. 133. See also the recent 'Post Office Horizon scandal' discussed in section 5.3. On the concept of 'safety-critical' and 'rights-critical,' see section 3 below.
[25] Don M. Gottfredson and Kelley B. Ballard, 'Association analysis, predictive attribute analysis, and parole behavior.' Paper presented at the Western Psychological Association meetings' (Portland: April 1964); Ernest W. Burgess, "Factors determining success or failure on parole" (1928) *The workings of the indeterminate sentence law and the parole system in Illinois*, 205–224.
[26] See literature cited above at n.13 and n.15.

*2.1    Statistical prediction models: conventional statistics vs machine learning approaches*

Statistical science is the mathematical study and use of data, including its collection, organization, analysis, interpretation and presentation, to better understand the world.[29] What we call a 'conventional' statistical approach is theory-driven, intended primarily to generate knowledge about an identified phenomenon by testing a pre-determined hypothesis (e.g., is there is a relationship between smoking and lung cancer?) to identify whether that hypothesis is correct (or at least plausible), thereby indicating that a stable, non-random relationship between those variables which can be generalised to a larger population. If such a relationship is found, this insight may be used to construct a statistical model to help predict future events and statistical values (e.g., what is the probability that a 40 year-old who has smoked regularly for 20 years will develop lung cancer?).[30] Accordingly, good statistical design requires domain-specific knowledge to generate a suitable hypothesis. Careful data selection is needed, both to ensure accurate representation of the relevant population so that subsequent analysis produces valid generalisations, and to avoid reducing the usefulness of a statistical model to inform understandings of the relationships between multiple variables. Hence, conventional statistical approaches to data analysis are best understood as *scientific endeavours* aimed at generating robust and replicable knowledge that proceeds through careful research design involving several steps: (1) formulation of a statistical hypothesis, (2) data collection (3) data analysis and testing of hypothesis (4) interpretation of results.[31]

ML techniques enable more advanced forms of statistical analysis via algorithms that utilise computational power to automate data analysis through the application of ML algorithms to historic datasets (which are often large and unwieldy in structure and organisation) to generate a statistical model.[32] To create a prediction model using ML, the data scientist does not begin with a pre-specified hypothesis but with an existing 'bag' of data, and then applies computational algorithms that automatically seek out patterns and correlations in the data to undertake the model-building process.[33] This helps explain the attraction of algorithmic tools for public authorities, particularly those with large troves of administrative data collected and retained over a sustained period yet held in a 'messy' fashion across disparate databases. Supervised ML techniques take a body of historic data and subject a portion ('training data') to analysis by general-purpose software algorithms that identify correlations within the data and construct a mathematical model that can generate predictions on unseen data.[34] Each training data point fed into the algorithm will contain multiple 'features', from which the trained

---

[27] See A. Ashworth and L. Zedner, *Preventive Justice* (Oxford: OUP, 2014), Ch.6; Harcourt, *Against Prediction: Profiling, Policing, and Punishing in an Actuarial Age;* Melissa Hamilton "Adventures in risk: predicting violent and sexual recidivism in sentencing law" [2015] Arizona State Law Journal 47(1), 11-62.
[28] For a more detailed account of how ML is used to build an algorithmic prediction model, see Part II of this paper.
[29] D. Spiegelhalter, *The Art of Statistics: Learning from Data* (Pelican, 2020), 1.
[30] Danilo Bzdok, Naomi Altman and Martin Krzywinski, "Statistics versus machine learning" (2018) Nature Methods 233-234;
[31] Agresti and Franklin, *Statistics: The Art and Science of Learning from Data*.
[32] Bzdok et al, "Statistics versus machine learning".
[33] Mannila "Data mining: machine learning, statistics, and databases."
[34] Peter Norvig and Stuart J. Russell, *Artificial Intelligence: A Modern Approach* (4th edn, New Jersey: Pearson Education, 2020). There are many different kinds of supervised learning algorithms. Some of the most common are linear regression, logistic regression, neural networks, naïve bayes, support vector machines, k-nearest neighbor, and random forests. At present, it appears that supervised learning techniques are most commonly used to build tools to inform public sector decisions, although in the absence of an authoritative inventory of these tools and their technical foundations, it is not

model will seek to identify correlations with the pre-specified phenomena to be predicted. This is an iterative, largely automated process involving continuous trial-and-error, and relies on 'learning' from feedback until a 'suitable' model is identified. In so doing, the data scientist makes important choices. Decisions about an 'acceptable' level of accuracy, for example, invariably depend upon the intended purpose of the tool - itself a product of organisational needs and the wider institutional context within which the algorithmic tool is embedded. As David Spiegelhalter explains:

> "Narrow AI refers to systems that can carry out closely prescribed tasks and there have been some extraordinary examples based on machine learning, which involves developing algorithms through statistical analysis of large sets of historical examples...Go, IBM Watson etc.... These systems did not begin by trying to encode human expertise and knowledge. They started with a vast number of examples, even by playing themselves as games. But again we should emphasise that these are technological systems that use past data to answer immediate practical questions, rather than scientific systems that seek to understand how they world works: *they are to be judged solely on how well they carry out the limited task at hand....*" (emphasis in original) [35]

ML techniques are now routinely used for online marketing and media content distribution by commercial organisations with the aim of boosting sales, click-through rates, user time-on-site and so forth, although, whether these techniques can be credited with bringing about such improvements remains unproven.[36] When used to inform organisational decision-making, algorithmic tools are best understood as a technology: a useful tool for achieving a given organisational purpose, rather than a scientific endeavour aimed at generating robust and replicable scientific knowledge about a given phenomenon.[37] As several management scholars explain:[38]

> "ML science had different goals from statistics. Whereas statistics emphasised being correct on average, ML did not require that. Instead, the goal was operational effectiveness. Predictions could have biases so long as they were better (something that was possible with powerful computers)...traditional statistical methods require the articulation of hypotheses or at least human intuition for model specification. Machine learning has less need to specify in advance what goes into the model and can accommodate the equivalent of much more complex models with many more interactions between variables."

Both conventional statistical and ML approaches operate through calculative processes that seek to identify correlations among data. These correlations are *not* indicative of causal relationships between data. But for many commercial applications, causal understanding is not necessary for these tools to be worth using. For example, if an ML-based ad-distribution model predicts that people who bought shampoo-X will also buy toothpaste-Z, automatically sending an ad for toothpaste-Z to shampoo-X buyers, neither high accuracy nor a causal explanation of the asserted relationship is necessary: if doing so coincides with increased sales of toothpaste-Z, that may be sufficient for the retailer to consider the automated service worthwhile, while mistaken predictions matter little to those receiving irrelevant product advertisements. Yet, as we shall subsequently argue in section 3.2 below, the underlying belief that 'only the patterns matter' prevalent in on-line content contexts should *not* be uncritically translated and applied to safety-critical or rights-critical domains.

*2.2    Understanding abstraction choices in algorithmic model-building*

We can better understand how computer scientists create and configure algorithmic tools to help inform (or even automate) real-world decisions, drawing on insightful cross-disciplinary work by legal scholar Andrew Selbst and his collaborators (hereafter 'Selbst et al').

---

possible to reach definitive conclusions about the nature of their use nor development. Other forms of machine learning techniques include unsupervised learning and semi-supervised learning.
[35] Spiegelhalter, *The Art of Statistics: Learning from Data*, 145.
[36] Mireille Hildebrandt, "The Issue of Proxies and Choice Architectures. Why EU law matters for recommender systems" (2021) https://osf.io/preprints/socarxiv/45x67/, 8, 11;.
[37] It is important to recognise, however, that ML *can* be and is used as a scientific tool for data-gathering for the purposes of scientific investigations aimed at producing robust, replicable knowledge.
[38] Ajay Agrawal, Avi Goldfarb, and Joshua Gans, *Prediction Machines: The Simple Economics of Artificial Intelligence* (Boston: Harvard Business Review Press, 2018), 40.

(a) Abstraction decisions and the 'contextual detachment' mindset of algorithmic tool-developers

To develop an algorithmic tool, the computer scientist must engage in 'problem framing' and 'goal identification.' This entails making various simplifying assumptions, or 'abstractions', seeking to remove the 'non-essential' domain-specific aspects of a problem ('noise') to create a mathematical model of a real-world task that can take input data and algorithmically convert it into outputs to inform that task. As Selbst et al explain:[39]

> "…when a computer scientist encounters a problem to solve, the first step is to identify what constitutes a set of well-defined inputs and secondly, well-defined outputs as a transformation of the input with desired properties: known in CS parlance as 'problem definition.' Having defined the problem, the next task is to identify the appropriate methods (or 'algorithms') to transform the inputs into the desired outputs as effectively as possible. Decisions about what constitutes an input and what constitutes a desirable output are the 'abstraction' choices made as part of the design of the resulting system, and they represent the boundary between the interior of the system and the exterior."

These abstraction choices play a critical role in computer science, particularly in machine learning, resting on crucial but rarely stated assumptions about the developer's role and responsibility. As Selbst et al further explain:

> "For the designer of the system, the processes that lead to the production of the inputs, as well as the processes that follow after the production are not to be considered part of the problem space and are disregarded. Correspondingly, for the (human) user of the system, the internal design choices used to construct the system are abstracted away: the user does not need to know how the input is transformed to the output as long as it does so correctly (i.e., in keeping with the specified properties of the output).[40]

> Hence issues such as 'fidelity of the training data set' and 'the appropriateness of the model for making decisions in the real-life task at hand' fall outside the computer scientist's abstraction boundary – and therefore not matters of concern or interest for which she considers herself responsible."[41]

These critical assumptions and abstraction choices are not, at present, typically documented or made explicit in a formal way, nor can inspection of an already-designed system reveal which abstraction choices were made and how.[42] Once made, these abstraction choices often fade into the background and become invisible, often rendering an explicit discussion about the design choices impossible.[43] Reliance on abstraction in the development of ML-based decision-support tools makes algorithmic models highly portable. As Scantamburlo, Charlesworth and Cristianini explain:

> "The principle underlying the use of machine decisions, that a score can be used as an indicator of the risk that a given person will behave in a certain way over a period of time (e.g., she will commit a crime, graduate successfully, or fulfil assigned tasks) remains the same for a broad range of uses, even if the consequences of a decision vary significantly in context."[44]

---

[39] Andrew Selbst, Suresh Venkatasubramanian and I. Elizabeth Jumar, "The legal construction of black boxes: How machine learning practice informs foreseeability" (2021) Paper presented at the We Robot 2021 conference, https://werobot2021.com/wp-content/uploads/2021/08/Kumar_et_al_Legal-Construction-of-Black-Boxes.pdf. (cited with the kind permission of the authors).
[40] Selbst et al., "The legal construction of black boxes."
[41] Selbst et al., "The legal construction of black boxes," 22.
[42] The EU's proposed Artificial Intelligence Act will require providers of 'high risk AI systems' (i.e., including ML-based decision-support tools deployed in criminal justice) to maintain 'risk management systems' (under Article 9), which includes a requirement to maintain, document and oversee 'relevant design choices' per Article 10(2)(a). Although we welcome the goals of improving transparency and accountability for ML-based tools, we doubt whether the Act will result in meaningful improvement due to its envisaged enforcement architecture: See Nathalie A. Smuha, Emma Ahmed-Rengers, Adam Harkens, Wenlong Li, James MacLaren, Riccardo Piselli and Karen Yeung, "How the EU can achieve legally trustworthy AI: A response to the European Commission's proposal for an artificial intelligence act" (2021) *https://papers.ssrn.com/sol3/papers.cfm?abstract_id=3899991*.
[43] Selbst et al., "The legal construction of black boxes: How machine learning practice informs foreseeability," 23.
[44] Teresa Scantamburlo, Andrew Charlesworth and Nello Cristianini, "Machine Decisions and Human Consequences" in K. Yeung and M. Lodge (eds.), *Algorithmic Regulation* (Oxford: Oxford University Press, 2019), 50-51.

However, because they effectively place boundaries on how the system might be expected to behave under 'normal' vs 'abnormal' conditions, including defining what 'normal' means, abstraction choices invariably have considerable impact, of varying seriousness depending upon the application context.[45] This conventional approach to problem-formation entails what we call a 'contextual detachment' mindset entailing the intentional 'detachment' or exclusion of contextual features from consideration during algorithmic model-building. However, as we shall subsequently demonstrate, matters that may be irrelevant 'noise' to the computer scientist can have important legal and constitutional implications if employed in criminal justice contexts.

(b) Socio-technical systems in constitutional context vs technical tools abstracted from reality

Because algorithmic tools used by organisations to facilitate efficient task-performance, rather than to discover scientific truths, evaluation of their 'quality' conventionally focuses on success in performing that task, typically understood in terms of prediction accuracy. For this purpose, model accuracy is typically understood a very narrow and specific sense, namely, accuracy in correctly predicting outcomes from a set of unseen historic data.[46] However, as Selbst and colleagues persuasively demonstrate, it is seriously deficient to focus these evaluations exclusively on the algorithmic model, rather than the larger socio-technical system in which they operate,[48] resulting in misleading and inappropriate 'quality' guarantees.[49] We argue, further, that algorithmic 'quality' also demands adherence to basic constitutional principles and legal requirements, at least for tools intended to inform public authority decisions. In these application domains, constitutional principles, administrative law doctrine and human rights norms are a crucial part of the relevant socio-technical context and must therefore inform and constrain the exercise of decision-making authority by criminal justice officials. They are not irrelevant 'noise.' To explain why, the following section proceeds in three parts. First, we explain why criminal justice contexts are 'rights-critical'. Secondly, we consider whether it is constitutionally legitimate for public authorities to use algorithmic prediction tools to inform how individuals are to be treated in the absence of a clear scientific understanding of the underlying causal relations between the variables upon which the prediction model relies. Thirdly, and primarily for the benefit of those unfamiliar with public law principles, we highlight three principles of particular importance in the context of criminal justice decision-making: the rule of law and the requirement of legality, respect for the basic liberty and human rights in general, and rights to due process in particular. We illustrate why these principles are particularly important by considering decisions by criminal justice authorities concerning whether to retain an arrested individual in custody.

## 3.  Constitutionalism: institutional safeguards against the abuse of public power

Although technical experts often describe criminal justice decisions as 'high stakes',[50] it is used to refer various circumstances in which the consequences of mistakes are likely to be very serious. To identify and ensure that appropriate safeguards are in place to prevent and protect against mistakes, we must clarify how and a given context is appropriately characterised as 'high stakes.'[51] For example, in engineering, it is well-established that 'safety-critical' contexts demand much more rigorous engineering design requirements and processes given the potentially devastating consequences of error and poor engineering design.[47] Engineers who design critical infrastructure recognise that they occupy positions of great power, demanding exceptional care in their design-decisions, and the need to ensure that construction faithfully follows these design-specifications. Similarly, medical advice provided by a clinician to his or her patient is also conventionally regarded as 'high stakes', for mistaken medical diagnostic or treatment decisions may have serious if not fatal consequences, prompting calls for meaningful and effective standardisation and oversight of ML-enabled tools to support medical decision-making.[48] Governmental office-holders, including but not limited to criminal justice authorities, also occupy positions of great power. But unlike structural engineers and medical professionals, their position of authority is derived not primarily from their professional expertise but is instead ultimately derived from the authority and trust placed in them by the community of citizens on whose behalf they are expected to serve, enabling them to engage in responsible governmental decision-making.[49]

---

[45] In the medical context, see Fedrico Cabitza, Davide Ciucci and Raffaele Rasoini, "A giant with feet of clay: on the validity of the data that feed machine learning in medicine" (2018) *https://arxiv.org/abs/1706.06838*.
[46] Discussed further in Part II, section 3.3.
[48] Selbst et al., "The legal construction of black boxes".
[49] Selbst et al., "The legal construction of black boxes".
[50] Michael Veale, Max Van Kleek and Reuben Binns, 'Fairness and accountability design needs for algorithmic support in high-stakes public sector decision-making' [2018] *Proceedings of the 2018 CHI Conference on Human Factors in Computing Systems* 440; Rudin, "Stop explaining black box machine learning models for high stakes decisions and use interpretable models instead."
[51] Brayne *Predict and Surveil*, 13: "…policing in practice means exercising an immense amount of discretion in the application of state power in high-stakes environments."
[47] Luiz Eduardo G. Martins and Tony Gorschek, "Requirements engineering for safety-critical systems: A systematic literature review" (2016) 75 *Information and Software Technology* 71-89.

In constitutional democracies, public officials possess a unique and distinctive power to undertake the task of governing, allowing them to make decisions and undertake actions that significantly affect the rights, interests and legitimate expectations of others. This powers carries with it onerous legal and constitutional duties and responsibilities. Within varied range of governmental functions undertaken by of public officials, the authority and position of criminal justice authorities is itself unique and distinctive, for they are empowered and duty-bound to enforce and uphold the criminal law, which entails conferring upon them extensive legal powers, including the power to punish criminal offenders. State-sanctioned punishment entails the lawful deprivation of basic rights and liberties of individuals in response to their commission of a criminal offence.[50] The form and severity of criminal punishment varies widely, ranging from long periods of imprisonment to the deprivation of property in the form of criminal fines, and is typically accompanied by significant moral and social stigma with serious consequences for the offender's life chances and opportunities.[51] But, unlike clinical or engineering design-errors, mistakes in criminal justice decision-making (with the exception of policing decisions concerning the use of force) are not typically 'safety critical' although their consequences may be are extremely serious. Instead, these consequences are best described as 'rights-critical', meaning that they may violate the legal and fundamental rights of affected individuals. In their most egregious form, mistaken criminal justice decision-making may result in the wrongful conviction and punishment of individuals for crimes they did not commit, producing serious injustice and devastating the lives of innocent persons and their families.[52] Furthermore, many criminal justice authorities, particularly the police, are invested with wide discretionary powers to enforce the law, and which are especially vulnerable to abuse, including powers to single out individuals considered to be 'of interest', subjecting them to heightened scrutiny, surveillance and interference, all of which are likely to be experienced as threatening, intrusive and unwelcome.[53]

Accordingly, modern constitutional democracies recognise that the power of the state, particularly the coercive authority of criminal justice officials, must be handled with great care,[54] reflected in their commitment to 'constitutionalism'.[55] Constitutionalism is rooted in recognition that the state's unique power, including its exclusive authority to establish and maintain the criminal justice system, is prone to abuse and overreach so that institutional mechanisms, including the need to maintaining a culture of vigilance and restraint, are essential and indispensable safeguards against the arbitrary exercise of power and the potential for despotism. Despotic rule—that is, the exercise of governmental power on the whim of a power-wielding despot,[56] is the antithesis of governance in accordance with clear, transparent, stable, prospective and generalised rules that are foundational to contemporary understandings of the rule of law.[57] Hence, constitutionalism insists that the exercise of government power should be open to public scrutiny (except in very limited circumstances requiring public justification) requiring those who wield governmental authority to bear legal and democratic responsibility for its exercise. Because the state's legitimate monopoly on coercive power underpinning the criminal law is invariably prone to abuse, the powers of criminal justice authorities within liberal democratic societies are necessarily limited and constrained by constitutional safeguards, including various public law principles. Although the precise meaning, scope and content of constitutional principles are often fiercely contested in legal debate, we portray them here in a simplified fashion to help those without legal training to understand their essential core. Before embarking on that account, a preliminary question arises concerning the constitutional legitimacy of using ML-based prediction models to make predictions about individuals without an understanding of the causal relationship between the input and output variables and to which we first turn.

---

[48] V. Sounderajah et al; STARD-AI Steering Committee. 'Developing a reporting guideline for artificial intelligence-centred diagnostic test accuracy studies: the STARD-AI protocol' (2012) 28 *BMJ Open* 11(6) e047709.
[49] A. Sajó and R. Uitz, *The Constitution of Freedom: An Introduction to Legal Constitutionalism* (OUP, 2017), 88-95; T. Endicott, *Administrative Law* (Oxford: OUP, 2011).
[50] H.L.A Hart, *Punishment and Responsibility* (Oxford: OUP, 2008).
[51] L. Campbell, A. Ashworth and M. Redmayne, *The Criminal Process* (5th edn., Oxford: OUP, 2019), 40.
[52] A.P. Simester and A. von Hirsch, *Crimes, Harms, and Wrongs: On the Principles of Criminalisation* (London: Hart, 2011), 4.
[53] Campbell et al., *The Criminal Process*.
[54] A. Ashworth and L. Zedner, *Preventive Justice*, 24.
[55] Sajó and Uitz, *The Constitution of Freedom: An Introduction to Legal Constitutionalism*.
[56] As Joseph Raz has put it, "What is arbitrary action generally? It is action indifferent to the reasons for or against taking it. Arbitrary government is the use of power that is indifferent to the proper reasons for which power should be used" in Joseph Raz, "The Law's Own Virtue" (2019) 39 *Oxford Journal of Legal Studies* 1–15, 5; Sajó and Uitz, *The Constitution of Freedom: An Introduction to Legal Constitutionalism*.

*3.1    Is it constitutionally acceptable for public authorities to use algorithmic prediction tools without understanding causality?*

Although this paper does not examine fundamental doubts about the epistemic validity and legitimacy of utilising statistical tools to inform criminal justice assessments of individuals, the turn to ML prediction models substantially enhances the danger of administering interventions based on either spurious correlations and/or on a misunderstanding of underlying causal pathways.[58] These dangers arise because the identified mathematical relationships between variables might be either**:**

(a) *purely coincidental*, i.e., based on 'spurious correlations' rather than a 'true' understanding of the relationship between the features identified by the ML model.[59] As already indicated, ML techniques operate by finding statistical correlations between the features and the labels associated with the training items, which are *not* indicative of causality. Although both conventional statistical methods and ML techniques can result in prediction models based on spurious correlations, this risk is much greater for ML models due to the 'curse of dimensionality'[60] and what statisticians call 'multiple testing'—that is, when testing multiple hypothesis simultaneously which are actually non-significant but can produce a significant result purely by chance;[61] or

(b) a reflection of 'true' real-world relationships between the features identified in the data, but those relationships are *wrongly interpreted* based on a mistaken understanding of their underlying causal mechanisms. The resulting correlations are 'true' (rather than spurious) but misinterpreted, resulting in mistaken inferences about the phenomena of interest. For example, an important study Caruana and colleagues discovered that a model trained to predict complications from pneumonia learned to associate asthma with a *reduced* risk of death.[62] Yet those with a fairly limited knowledge of asthma and pneumonia recognised that this was obviously wrong.[63] The model was trained on clinical data from past pneumonia patients. Patients who suffer from asthma truly *did* end up with better outcomes, so the model was not built on spurious correlations, but on true relationships in the data. However, the model failed to recognise that, because asthma sufferers regularly monitor their breathing, they go to hospital *earlier* than other pneumonia patients and, on arrival at hospital, received more immediate and focused treatment because they are considered higher risk. Hence, their treatment resulted in better outcomes than pneumonia patients who were not asthma sufferers.[64]

---

[57] J. Raz, "The rule of Law and its Virtue" in *The Authority of Law: Essays on Law and Morality* (Oxford: OUP, 1979); Paul Craig, "Formal and substantive conceptions of the rule of law: an analytical framework" [1997] *Public Law* 467-487.
[58] See Hamilton "Adventures in risk: predicting violent and sexual recidivism in sentencing law."
[59] Haig explains that the term 'spurious correlation' is ambiguous in the methodological literature. Sometimes it refers to situations in which, in a system of three variables, the existence of a misleading correlation between the two variables is produced through the operation of a third variable. At other times it refers to 'nonsense' correlations, which are accidental correlations for which no sensible natural causal interpretation can be provided, and which we adopt here. See Brian D. Haig, "What is a spurious correlation?" (2010) 2(2) *Understanding Statistics: Statistical Issues in Psychology, Education, and the Social Sciences*, 125-132.
[60] The curse of dimensionality arises because it becomes more difficult to distinguish true predictive capacities of a model from luck as the volume of data and number of variables increases: C. T. Bergstrom, J. D. West, *Calling Bullshit: The Art of Skepticism in a Data-Driven World* (New York: Random House, 2020), 475.
[61] Scantamburlo et al., "Machine Decisions and Human Consequences," 56.
[62] Rich Caruana, Yin Lou, Johannes Gehrke, Paul Koch, Marc Sturm and Noemie Elhadad, "Intelligible models for healthcare: Predicting pneumonia risk and hospital 30-day readmission" (August 2015) *KDD '15: Proceedings of the 21st ACM SIGKDD International Conference on Knowledge Discovery and Data Mining* 1721-1730, 1721.
[63] Caruana et al., "Intelligible models for healthcare".

In either case, the use of ML prediction tools to inform organisational decisions in safety-critical and rights-critical contexts may produce very serious consequences. If these predictions inform medical diagnosis, treatment or even triaging decisions, for example, mistakes may well be fatal. Similarly, if the substantive interventions that are informed by these predictions are rights-critical, including decisions by criminal justice authorities concerning the treatment of individuals, they may produce serious injustice. Yet in all cases, the predictions are likely to be cloaked with the sheen of objectivity frequently associated with the use of predictive analytics, while both developers and the organisations commissioning them fail to recognise that their predictions and the models generating them are best understood as 'bullshit.'[65] In short, an important yet often overlooked question concerns identifying the conditions under which algorithmic prediction tools, however created, can legitimately inform real-world decisions in the absence of a clear scientific understanding of the underlying causal relations between the variables upon which the model relies. While organisations that deploy ML-based tools for on-line retailing, marketing, and commercial customer relationship management may not care *why* particular correlations arise repeatedly, the underlying belief that 'only the patterns matter' should *not* be uncritically translated and applied to safety-critical or rights-critical domains.

These concerns are related to demands for algorithmic 'explainability'[66] due to the opacity of algorithms and our limited ability to understand and comprehend how their outputs are generated.[67] A burgeoning field of technical research in 'explainable AI' (XAI) seeks to provide visibility into how deep learning systems that are largely uninterpretable to humans (sometimes referred to as 'black box' models) produce outcomes and predictions.[68] However, the functional 'explanations' which computer scientists are concerned with should *not* be confused with the need for reasoned justification, particularly for algorithmic tools used to inform public authority decision-making. Even if adequate functional explanations can be produced, that is not in itself a normative justification for using it to inform any particular decision, particularly those made by criminal justice authorities about individuals. As Selbst and Barocas argue:

> "…that decisions based on machine learning reflect the particular patterns in the training data cannot be a sufficient explanation for why a decision is made the way it is."[69]

Nor does a functional explanation overcome the lack of understanding of the causal relationship between the input variables and the phenomenon of interest the model claims to predict. Hence some scholars make the related claim that ML models must themselves be 'interpretable' so humans can readily understand how an output was generated from a given set of inputs, at least for 'high stakes' decision-making.[70] As Selbst and Barocas observe, this appeal to model-interpretability rests on a belief in the value of identifying a plausible 'story' or narrative interpretation of the data in which human intuition is typically employed as a bridge from explanation to normative assessment. They acknowledge that intuition offers a powerful, ready mechanism through which humans can draw on a wealth of accumulated insight and experiences to evaluate ML models to discount patterns that violate well-honed expectations and to recognize and affirm discoveries that align with experience. Moreover, domain experts can provide this 'common sense' conception of intuition, strengthening the capacity to see where models may have gone awry and reflecting long-standing reliance by social scientists on 'face validity' to evaluate whether a model is measuring what it purports to measure. But although intuition is useful, they warn that intuition can be wrong. It can lead to the rejection of valid models because they are unexpected or unfamiliar, or to endorse false discoveries because they align with existing beliefs, fostering 'just so' stories that appear to make good sense of the facts as presented but are actually unreliable.[71] For them, intuition is *less useful* when dealing with findings that fail to comport with or even run counter to experience. Moreover, the entire 'value proposition' underpinning the use of ML techniques lies in the capacity to generate insights by finding patterns in large data-sets that extend well beyond human intuition. Hence ML models might depart from intuition in ways that reflect the 'true' underlying relationships between variables, yet do not lend themselves to hypotheses that can readily account for the model's discoveries.[72]

---

[64] Caruana et al., "Intelligible models for healthcare".
[65] Bergstrom, J. D. West, *Calling Bullshit: The Art of Skepticism in a Data-Driven World*. Frederike Kaltheneur, *Fake AI* (Meatspace Press, 2021); Arvind Narayanan. "How to recognize AI snake oil." *Arthur Miller Lecture on Science and Ethics* (2019).
[66] Legal scholars (and others) have been particularly pre-occupied with the 'explainability' requirements arising under the GDPR: see Williams "Rethinking administrative law" 474-476 (and literature cited therein).
[67] Jenna Burrell, "How the machine "thinks": Understanding opacity in machine learning algorithms" (2016) *Big Data & Society* 1-12.
[68] Rudin, "Stop explaining black box machine learning models for high stakes decisions and use interpretable models instead."
[69] Andrew D. Selbst and Solon Barocas, "The intuitive appeal of explainable machines" (2018) 87 *Fordham Law Review* 1085-1139, 1126. See also footnotes 259 and 260 in same text: reliance on measurements of accuracy alone is also not a sufficient normative justification for the use of a machine learning-based tool that uses a non-interpretable model.
[70] See discussion at beginning of section 3 concerning variety across different 'high stakes' contexts.

In other words, to insist upon scientific proof of the underlying causal relationships may be too demanding, given difficulties associated with establishing the true 'effects' of real-world interventions due to the impossibility of isolating specific causes and establishing counter-factual events. It might rule out decision-support tools that are genuinely valuable, particularly when based on 'true' correlations between data features even if the underlying causal pathways remain uncertain. It is beyond the scope of this paper to develop a comprehensive theory to identify when normative justifications of ML decision-making models are needed to lawfully or legitimately utilise these models. Nonetheless, we suggest that that any acceptable framework for analysis for identifying when public authorities (including, but not limited to criminal justice authorities) can be justifiably informed by data-driven predictions, at least five matters must be considered:

i. First, a threshold question arises about whether normative justifications are needed to support the proposed intervention. Only if the intervention adversely affect the rights, interests, and legitimate expectations and/or redistributes resources, risks, or benefits across the community in a significant and material way do liberal constitutional democracies generally require public authorities to provide reasoned justifications for such actions.[73] This clearly includes predictions used by criminal justice authorities to inform the treatment of individuals but might not, for example, extend to the fonts and colours of a public authority website;[74]

ii. Secondly, if normative justifications are required, this extends to the tools used to inform those decisions. Within liberal democratic societies, public authority decisions must be normatively justified for two reasons: firstly, because their power is ultimately derived from the citizens on whose behalf public authorities are required to serve[75] and to whom they are constitutionally obliged to account—and secondly, due to the serious dangers associated with the abuse of governmental power;[76]

iii. Thirdly, in the absence of scientific evidence to support the model's predictions, some plausible hypotheses about the relationship between the relevant features and the prediction generated by the model should be publicly stated, supported by additional procedural safeguards to protect against the dangers of errors and resulting injustice. This includes a right to challenge the system's predictions and a right to rebut the claimed inferences associated with those predictions. For example, in a challenge to the ML-based decision-support system called SyRI used by several Dutch public authorities to help detect and prevent tax and social security fraud,[77] reference was made to the 'Waterproof project' intended to identify 'living together' benefit fraud (where an individual falsely claims to be living alone to receive a higher benefit-rate). Annual water usage data from water bills of individual residents were matched against welfare authority data to identify those with particularly low water usage levels and thus deemed 'at risk' of committing benefit fraud. While a plausible explanation for the posited relationship between low water usage and living together benefit fraud was offered, individuals highlighted as 'risky' on this basis should be routinely entitled to explain why their water bills are low thereby rebutting the asserted logic of the system. For example, it was later found that 20% of those flagged as risky were erroneously singled out because they had valid reasons for low recorded water usage, including broken water meters and the use of household water recycling systems;[78]

---

[71] Selbst and Barocas, "The intuitive appeal of explainable machines," 1129.
[72] Selbst and Barocas, "The intuitive appeal of explainable machines," 1129.
[73] cf R. Forst, *The Right to Justification* (New York: Columbia University Press, 2011); Rainer Forst, "The Justification of Human Rights and the Basic Right to Justification: A Reflexive Approach" (2010) 120 *Ethics* 711, 734; D. Kyritsis, *Where Our Protection Lies* (Oxford, OUP, 2017), 66.
[74] However, the use of online behavioural profiling and manipulation has demonstrated that manipulative interventions can be achieved via subtle alterations to digital interfaces, such as font size, style and colour, and are used to manipulate consumers and voters into behaviours preferred by vendors or political actors, without their knowledge ('dark patterns'). See, for example, Jamie Liguri and Lior Jacob Stahilevitz, "Shining a light on dark patterns" [2021] 13(1) *Journal of Legal Analysis* 43-109.
[75] For example, F. Wendt, "Rescuing Public Justification from Public Reason Liberalism" in D. Sobel, P. Vallentyne and S. Wall (eds.) "Oxford Studies in Political Philosophy Volume 5" (Oxford: OUP, 2019) and literature cited therein.
[76] See section 3 above.

iv. Fourthly, even if a plausible hypothesis about the relationship between relevant data features and resulting prediction is offered (such as low water-meter bills indicating welfare fraud), it does not follow that the underlying data can be legitimately or lawfully used to generate adverse predictions of this kind.[79] If these predictions are generated by subjecting data to algorithmic processing for purposes that violate an individual's reasonable expectations of privacy (based on a theory of contextual integrity),[80] the resulting privacy interference can only be justified if the public authority can demonstrate that using the data for that purpose is necessary and proportionate in a democratic society to meet a pressing social need.[81] In the Waterproof example, Dutch authorities subjected personal data from the water bills of 63,000 persons to algorithmic analysis, resulting in 457 flagged as 'at risk' but ultimately only 42 (i.e., 0.07%) found to have committed fraud.[82] It was hardly surprising that the court concluded that the burden of justifying the scale and seriousness of the interferences of the Art 8(1) ECHR right of affected persons was not established. Moreover, although such crimes are serious insofar as they are committed culpably and with dishonest intent, they are relatively minor in terms of damage and scale of harm, involving no violence to persons or damage to property;

v. Fifthly, the constitutional acceptability of using an algorithmic predictions to inform how individuals are treated by the state depends critically upon the nature and significance of the substantive intervention that the output is intended to inform. In a celebrated example of a decision by a public authority that would be set aside as unlawful in English administrative law on the basis of 'irrationality' (the so-called 'Wednesbury unreasonableness' test ), Lord Justice Warrington in *Short v Poole Corp* [1926] Ch 66, 90-91 referred to the hypothetical dismissal of a red-haired teacher because she had red hair as a decision as "so absurd that no reasonable person could dream that it lay within the powers of the decision-making authority and thus would be invalidated by a court on an application for judicial review". Although this example is typically lamented due to the dubious role of hair colour as an indicator of professional competence, what makes the example so egregious is that hair-colour is used as the basis for terminating an individual's contract of employment.[83] If, however, a public health authority uses red-hair colour to identify proposed recipients of health advice encouraging them to avoid the sun during hot weather (given that red-haired persons hair carry mutations in the gene MC1R and thus at far greater risk for skin melanoma that those without this gene variant[84]) then this is unlikely to be constitutionally problematic. Not only would the proposed intervention informed by a proper causal understanding of the relationship between the input variable (hair colour) and output recommendation (risk of skin melanoma), but it would not adversely impact the recipient's rights, interests or legitimate expectations—they would be merely offered health advice which they are free to ignore.

---

[77] NJCM et al. and FNV, Rechtbank den Haag, ECLI: NL: RBDHA: 2020:1878 ("SyRI Judgment").

[78] SyRI Judgment at [3.1]-[3.3].

[79] For a more complete discussion of the unlawful collection and processing of personal data, see section 3.1(a) of Part II of this paper.

[80] Helen Nissenbaum, "Privacy as contextual integrity" (2004) 79 *Wash. L. Rev.* 119.

[81] European Convention on Human Rights ("ECHR"), Art. 8(2)

[82] SyRI Judgment at [3.1]-[3.3].

[83] T. Endicott, "The value of vagueness" in V.K. Bhatia, J. Engberg, M. Gotti and D. Heller (eds.) *Vagueness in Normative Texts* (Bern: Peter Lang, 2005), 27-48.

[84] Sara Raimondi, Francesco Sera, Sara Gandini, Simona Iodice, Saverio Caini, Patrick Maisonneuve and Maria Cocetta Fargnoli, "MC1R variants, melanoma and red hair color phenotype: A meta-analysis" (2008) 122(12) *Cancer Genetics* 2753-2760.

*3.2 Constitutional safeguards against arbitrariness and injustice in criminal justice decision-making: the rule of law, respect for human rights and rights to due process and fair procedure*

Even if the decision to adopt an algorithmic tool to inform decisions by criminal justice authorities about the treatment of individuals can be adequately justified in the face of possible constitutional objections, public law principles also ground more specific legal obligations that public authorities must comply with when exercising discretionary power. Accordingly, decision-support tools used to inform the exercise of that decision-making power must also comply with them. Although the range of specific legal duties will vary between jurisdictions, they are likely to include duties arising under administrative law, and – more recently, legal duties to ensure respect for human rights, compliance with data protection laws and equalities legislation. Although a number of constitutional principles spring from a commitment to constitutionalism including, for example, the separation of powers and the independence of the judiciary, three are particularly important in criminal justice contexts: the rule of law and the requirement of legality, respect for the basic liberty and human rights in general, and rights to due process in particular. The following discussion briefly explains, primarily for the benefit of the non-lawyers, the foundational core of each, illustrating why they are particularly important by reference to decisions by criminal justice authorities concerning whether to retain an arrested individual in custody.

(a)     The rule of law and the need for legal authorisation of state action

Central to the rule of law is the requirement of 'equality before the law', meaning that no person or body, including the state itself, is above the law.[85] It requires that the state must comply with the general laws that apply to ordinary persons, such as health and safety laws, data protection laws, equality laws, and so forth, unless specific legal exemptions apply. Importantly, within liberal democratic states, the rule of law imposes obligations on the state that are considerably more onerous than those imposed on ordinary persons. Unlike ordinary persons, who are free to act as they wish provided they do not violate the law,[86] all state activity must be 'authorised by law' so that if a decision made by the state exceeds its legal authority, it will be unlawful and invalid, or liable to be invalidated.[87] This protects individuals and the public from dangers that would arise if those in government were free to arrogate unlimited power to themselves in pursuit of purposes, and or through means and processes, that have not been legally authorised.

(b)     The basic liberty of the individual and respect for human rights

The need for legal authorisation of state action reflects an important difference between the legal position of the individual and that of the constitutional democratic state. While individuals are free to live according to their own view of the good provided they abide by law, the authority of the liberal democratic state is ultimately rooted in the consent of its citizens, on whose behalf the state governs and is expected to serve. Public authorities are also duty-bound to respect human rights. In the UK, for example, the Human Rights Act 1998 ('HRA') mandates compliance with the rights set out in the European Convention on Human Rights ('ECHR') such that, except in relation to rights which admit of no qualification, rights can be restricted only in very limited circumstances where necessary and proportionate in a democratic society to meet a legitimate and pressing social need.[88] For example, although the basic right to liberty protected by Article 5(1) ECHR provides that 'everyone has the right to liberty and security of the person', the practice of imprisonment (coercively detaining individuals in custody) is considered a legitimate interference with this right if in the form of 'lawful arrest' or 'lawful detention', and provided the conditions of Article 5(1) are met.[89] Such rights provide critical safeguards against the imprisoning of dissidents and other undesirables on the arbitrary say-so of governmental officials, a practice central to the playbook of authoritarian and repressive governmental regimes throughout history.

---

[85] Paul Craig, "Formal and substantive conceptions of the rule of law: an analytical framework," 4; Raz, "The rule of Law and its Virtue," 224. Although there is widespread agreement about the 'core' commitments of the rule of law, its precise content and contours are subject to intense on-going contestation: e.g.; D. Dyzenhaus, 'Recrafting the Rule of Law' in David Dyzenhaus (ed.), *Recrafting the Rule of Law: The Limits of Legal Order* (London: Hart, 1999);.
[86] *The Case of Proclamations* [1610] EWHC KB J22: "the King hath no prerogative, but that which the law of the land allows him."
[87] D. Galligan, *Law in Modern Society* (OUP, 2006), 92-94.
[88] See for example the Right to Private and Family Life under European Convention on Human Rights ("ECHR"), Article 8(2) .
[89] See Article 5(1) ECHR.

(c)     Procedural fairness and rights of due process

The rule of law also includes the principle of 'no punishment without law': no person can be lawfully punished unless and until *convicted* for committing a criminal offence.[90] The law confers systematic procedural safeguards on those accused of a criminal offence, including rights to receive timely and informative notice of a hearing, to know the case against them, to a fair and impartial hearing, to have an opportunity to respond, to access legal counsel, to published reasons for the decision, alongside other procedural rights including, in some cases, an ability to appeal the decision or seek judicial review. While the European tradition historically refers to these rights under the rubric of 'equality of arms', many common law and civil law jurisdictions refer to this bundle of normative legal rights and presumptions as rights of 'due process' or 'procedural fairness.'[91] These legal safeguards protect the individual against the grave moral injustice that arises from the conviction and punishment of the innocent, grounding the 'right to a fair trial' long recognised as fundamental to the fair administration of justice. In contemporary form, Article 6 ECHR accords more specific rights to those charged with a criminal offence, including the presumption of innocence (Article 6(2)) and the right to be promptly informed of the nature and cause of the accusation against him (Article 6(3)(a)). More specific rights accrue to those arrested by state authorities under Article 5(2) ECHR, which requires (among other things) that 'everyone who is arrested shall be informed promptly, in a language which he understands, of the reasons for his arrest and of any charge against him.' National jurisdictions typically confer more specific procedural rights upon individuals, varying in their content and contours. For example, the right to procedural fairness recognised in British administrative law demands that the decisions of public authorities (1) must be free from bias, which, in the case of judicial and other adjudicative decisions, includes both actual bias and the appearance of bias, and (2) an individual whose rights, interests or legitimate expectations are adversely affected by a decision are offered a meaningful right to participate in that decision (often described as the 'right to be heard'), including the right to be informed about a decision and to contest it before an impartial adjudicator.[92]

*3.3    An illustration: arrest, detention and 'mistaken' decisions*

To illustrate how the preceding three constitutional principles described above are operationalised as specific legal safeguards that constrain criminal justice decision-makers while protecting individuals against unjustified violations of their rights, consider the position of an individual (D) who is arrested by a police officer and taken into police custody, thereby depriving her of liberty of movement. To effect a lawful arrest in England and Wales, the conditions set out in the Police and Criminal Evidence Act 1984 ("PACE"), s.24 and PACE Code G must be complied with. These require, among other things, that the arresting constable has 'reasonable grounds' to suspect that arrestee has, is in the process of, or is about to, commit a criminal offence and secondly, for believing that the person's arrest is 'necessary' for certain specified purposes set out in s.24(4)-(5). PACE also imposes a series of legal duties on the arresting officer, including requirements that the officer create an arrest record identifying the grounds for the arrest, the reasons why the arrest was necessary, confirmation that the arrestee was formally cautioned prior to arrest (unless excepted), informed of his or her right to silence, and anything said by the arrestee at the time of arrest.[93] The arrestee must also be informed at the time of arrest (or as soon as possible thereafter), that they are under arrest and the grounds and reasons for their arrest and can normally only be detained for up to 24 hours without charge.[94] All of these legal duties can be readily understood as grounded in the three constitutional principles identified above. Thus, the powers of police officers to arrest an individual stem primarily from the grant of lawful authority established under PACE, which requires arrests to be conducted in accordance with PACE, otherwise they will be unlawful. Likewise, the duties to record material details of the arrest, to inform the arrestee of her rights, and time limitations imposed on the maximum duration of detention, are all rooted in the individual's right to due process and the protection of the arrestee's basic right to liberty.

---

[90] A.V. Dicey, *Introduction to the Study of the Law of the Constitution* (8th edn., Liberty, 1982), 187. See also Article 5(1) ECHR *ibid*.
[91] I. Kerr, 'Prediction, pre-emption, presumption: the path of law after the computational turn' in M. Hildebrandt and K. de Vries, (eds.), *Privacy and Due Process After the Computational Turn*, (Oxford: Routledge, 2013), 108.
[92] D. Galligan, *Due Process and Fair Procedures: A Study of Administrative Procedures* (Oxford: OUP, 1996). Endicott, *Administrative Law*, Chapter 4 and cases cited therein; On relevant human rights cases, see B. Dickson, *Human Rights and the United Kingdom Supreme Court* (OUP, 2013), and cases cited therein.
[93] PACE Codes of Practice Code G, s 4.1
[94] PACE, s.41.

At the end of the permitted arrest period[95], decisions must be made about whether to charge D, and if charged, whether to release D on bail. If charged with a criminal offence, D may be released on bail (which may include bail conditions). If bail is denied, this entails a serious interference with D's liberty, often for a very substantial period, with highly consequential implications for D and D's family while awaiting trial.[96] Accordingly, the decision to deny bail can only be made if the conditions of s.3(6) of the Bail Act have been met which include (among other things) that detention is deemed 'necessary' to secure D from absconding, to prevent D committing an offence while on bail, or to protect D or another person.[97] It is in making determinations of this kind that criminal justice authorities in US states have enthusiastically embraced algorithmic decision-support tools that purport to assess the risk that an individual will abscond ('flight risk'), or commit a criminal offence if released,[98] although we are not aware of their use by British courts for these specific purposes. 'Mistaken' assessments concerning whether it is 'necessary' to remand D to prevent such occurrences are 'rights-critical' if it results in the coercive detention of a person who is in fact unlikely to abscond (so-called 'false positives'), commit a criminal offence or cause harm to another had they been released on bail.[99] On the other hand, mistakes that flow in the opposite direction (so-called 'false negatives') – in which a determination is made that remand in custody is not necessary to prevent that person committing a criminal offence, or to protect another, for example – may result in public release of an individual who is in fact likely to commit a violent offence, posing serious threats for vulnerable individuals, particularly victims of domestic violence.[100] Although resolving the tension between the need to respect the rights of those who have been charged but not yet convicted and legitimate concerns to protect public security generally, and vulnerable victims in particular, are typically matters requiring case-by-case judgement by a magistrate, the use of algorithmic decision-making support tools effectively encodes a predetermined assessment of the desired balance into a mathematical prediction model.[101] It is precisely these sorts of 'technical' decisions that we examine in more detail in Part II demonstrating why such decisions have direct and important constitutional implications that are reflected in jurisdiction-specific legal duties and obligations and *cannot* be satisfactorily resolved solely by reference to computational know-how.

## 4. Conclusion

This Part I paper has shown how the use supervised ML techniques to build algorithmic prediction models for use in decision-support tools conventionally entails abstraction decisions such that the social context into which algorithmic predictions are intended to work are 'detached' from the model-building process, without reference to larger socio-technical context in which the model works. Accordingly, constitutional safeguards intended to guard against the dangers arising from arbitrary decision-making that apply to public authorities, including 'rights-critical' decisions that criminal justice authorities are frequently required to make, are typically ignored in the algorithmic model-building process. As a result, decisions that rely on algorithmic predictions are being made without the protection offered by constitutional safeguards and hence more likely to result in arbitrary decision-making, producing injustice. In Part II, we pinpoint more precisely how seemingly 'technical' choices made by developers when building these tools have serious constitutional and legal implications by reference to three algorithmic decision-tools either in use or recently in use for criminal justice purposes, and reflect on their implications for the practice of algorithmic model-building and implementation for public sector use and the need for more systematic, effective and practical constitutional safeguards.

---

[95] Normally 24 hours unless an extension has been granted under the Policing and Crime Act 2017, s.62, and PACE, s.30.
[96] As of December 2021, the total population of individuals held on remand stood at 12,780 (a 6% increase from 2020, due to court delays caused by the Covid-19 pandemic). The median number of days spent on remand by individuals due to be tried at the Crown Court stood at 449 days in Autumn 2021 (an 11% increase from 2020).
[97] Bail Act 1976, s.3(6). The decision whether to refuse bail and to remand D pending trial is made by a magistrate.
[98] Kathrin Hartmann & Georg Wenzelburger, "Uncertainty, risk and the use of algorithms in policy decisions: a case study on criminal justice in the USA" (2021) 54 *Policy Sciences* 269-287.
[99] But see A. Hucklersbury, "Bail or Jail? The Practical Operation of the Bail Act 1976" (1996) 23(2) *Law and Society Review* 213-233.
[100] Following the murder of Kay Richardson her ex-partner following his release under investigation, despite evidence of previous domestic abuse, new laws and statutory guidance reforming pre-charge bail to provide stronger protection to victims have been proposed: Home Office, "Home Secretary to introduce 'Kay's Law' reform to better protect victims" (14 January 2021) *https://www.gov.uk/government/news/home-secretary-to-introduce-kays-law-reform-to-better-protect-victims*; "Home Office response to recommendations from 'A duty to protect'" (7 June 2022) *https://www.gov.uk/government/publications/responses-to-super-complaint-report-a-duty-to-protect/home-office-response-to-recommendations-from-a-duty-to-protect-accessible*, Recommendation 4.
[101] See section 3.2. of Part II of this paper, where we discuss the trade-off between Type I and Type II errors in configuring error thresholds of algorithmic tools.

**5.12.22**
11933 words